\documentclass[pre,showpacs,preprint]{revtex4}

\usepackage{amsmath}
\usepackage{color}
\usepackage{graphicx}
\usepackage{dcolumn}
\usepackage{bm}

\begin{document}

\title{Uniform self-diffusion in a granular gas}
\author{J. Javier Brey and M.J. Ruiz-Montero}
\affiliation{F\'{\i}sica Te\'{o}rica, Universidad de Sevilla,
Apartado de Correos 1065, E-41080, Sevilla, Spain}
\date{\today }

\begin{abstract}
A granular gas composed  of inelastic hard spheres or disks in the homogeneous cooling state is considered. Some of the particles are labeled and their number density exhibits a time-independent linear profile along a given direction. As a consequence, there is a uniform flux of labeled particles in that direction. It is shown that the inelastic Boltzmann-Enskog  kinetic equation has a solution describing this self-diffusion state. Approximate expressions for the transport equation and the distribution function of labeled particles are derived. The theoretical predictions are compared with simulation results obtained using the direct Monte Carlo method to generate solutions of the kinetic equation. A fairly good agreement is found.

\end{abstract}

%\pacs{05.20.Dd,51.10.+y,05.60.-k}

\maketitle

\section{Introduction}
\label{s1}
The description of granular gases in terms of a more fundamental underlying kinetic theory has been developed with considerable success \cite{Go03,ByP04}. Traditional methods of nonequilibrium statistical mechanics have been adapted to the peculiarities of model systems of granular gases, mainly the inelasticity of collisions implying the absence of an equilibrium state. In some cases, granular gases stretch kinetic theory and also hydrodynamics to their limit \cite{Go99}.

A prototype of transport process is self-diffusion, i.e. the diffusion of a two-component fluid when the particles of both components are mechanically equivalent. For elastic systems, this process has been extensively studied and the conditions for a macroscopic diffusion equation and the form of the latter have been identified with quite generality \cite{DyvB77,McL89}. Self-diffusion in granular gases has also been studied \cite{BRCyG00,ByP00,DByL02,LByD02}. In all the latter studies, and also in most of those carried out in molecular fluids, the limit of  very low concentration of one of the components (tracer limit) is considered, and the aim is to derive the self-diffusion hydrodynamic equation for that component to Navier-Stokes order, i.e. to get an expression for the particle flux that is valid to first order in the density gradient.

In this paper, self-diffusion will be considered in a situation in which there can be a macroscopic flux of labeled particles in the system, since the concentrations of both components are of the same order. Moreover, nothing will be assumed on the magnitude of the  concentration gradients, so that the validity of the study is not restricted to the small gradient limit. Finally, the state considered will be characterized by both a stationary density profile of tagged particles and a uniform flux of them. This state will be referred to as the uniform self-diffusion state. The study presented here has been prompted and stimulated by the analysis of a similar one in the context of molecular, elastic fluids, which are globaly at equilibrium \cite{ByR13a}. Of course, there are significant differences to confront. Instead of an equilibrium situation, the homogeneous cooling state (HCS) of a granular gas is considered. The distribution function of this state has the property that all its time dependence is entirely through the mean-square kinetic energy, or granular temperature, of the particles. This scaling property turns out to be crucial both for the theoretical study of the uniform self-diffusion state and also to carry out particle simulations of the state in an efficient way. Besides, the HCS of a granular gas is known to be unstable against perturbations of large enough wave-length \cite{GyZ93,BRyM98}.

In spite of the differences with the elastic case, it will be shown that the Boltzmann-Enskog equation for inelastic hard spheres or disks admits a solution having the macroscopic properties characterizing the uniform self-diffusion state, in a system that is globally in the HCS. Although the exact form of the solution is not known, the transport equation, the expression of the involved transport coefficient, and even the one-particle distribution function of labeled particles, will be obtained by means of an approximation.  The theoretical predictions will be validated by comparing with numerical results obtained by the direct simulation Monte Carlo method. To generate the desired state, non-local boundary conditions will be employed, in the spirit of non-equilibrium particle  simulation methods.

It is worth to emphasize that solutions of the Boltzmann-Enskog equation describing non-equilibrium states with arbitrarily large gradients of hydrodynamic fields are scarce, but they provide a solid and unique starting point to investigate macroscopic properties of non-equilibrium transport.

The remaining of the paper is organized as follows. In the next section, some properties of the inelastic Enskog equation as applied to the HCS are  reviewed briefly. The macroscopic state of uniform self-diffusion being considered is defined in Sec.\ \ref{s3}, where the Enskog equation obeyed by the one-particle distribution of labeled particles is presented. Then, an approximated  solution describing the macroscopic state of uniform self-diffusion for arbitrary concentration and density gradient of labeled particles is obtained. The approximation scheme used is based on a property of the linearized kinetic equation, rather than on an expansion in orthogonal polynomials, as it is usually done. Nevertheless, the results are quantitatively close to those derived by using a  truncated Sonine polynomial expansion. The theoretical predictions are compared with numerical solutions of the Boltzmann equation, constructed by means of the direct simulation Monte Carlo method, and a good agreement is found. The comparison includes not only the hydrodynamic profiles and the transport coefficient, but also the one-particle distribution of labeled particles. The last section of the paper contains a short summary and some additional comments on the results derived.

\section{The inelastic Enskog equation and the homogeneous cooling state}
\label{s2}
The system under consideration is a granular gas composed of $N$ equal hard spheres ($d=3$) or disks ($d=2$) of mass $m$ and diameter $\sigma$. Collisions between particles are inelastic, and characterized by a constant, velocity independent, coefficient of normal restitution $\alpha$, which is defined in the interval $0 < \alpha \leq 1$. It is assumed that the behaviour of the system is accurately described by the (inelastic) Enskog equation. Then, the one-particle distribution function $f({\bm r},{\bm v},t)$ giving the average number of particles at position ${\bm r}$ with velocity ${\bm v}$ at time $t$ satisfies the equation \cite{BDyS97}
\begin{equation}
\label{2.1}
\left( \frac{\partial}{\partial t} + {\bm v} \cdot \frac{\partial}{\partial {\bm r}} \right) f({\bm r}, {\bm v},t)= J_{E}[{\bm r},{\bm v}|f,f],
\end{equation}
where the inelastic Enskog collision operator $J_{E}$ is defined as
\begin{equation}
\label{2.2}
J_{E}[{\bm r}_{1},{\bm v}_{1}|f_{1},f_{2}]= \int d{\bm r}_{2} \int d{\bm v}_{2}\, \overline{T}_{+}(1,2) f_{1}({\bm r}_{1},{\bm v}_{1},t) f_{2}({\bm r}_{2},{\bm v}_{2},t) g_{E}({\bm r}_{1},{\bm r}_2,t),
\end{equation}
for arbitrary functions $f_{1}({\bm v})$ and $f_{2} ({\bm v})$. Here $g_{E} ({\bm r}_{1},{\bm r}_{2},t)$ is the pair correlation function of the fluid. In the revised Enskog theory (RET) \cite{vByE79}, it is approximated by the equilibrium functional of the density evaluated for the local density field of the fluid. The binary collision operator $\overline{T}_{+}(1,2)$ is
\begin{equation}
\label{2,3}
\overline{T}_{+}(1,2) = \sigma^{d-1} \int d \widehat{\bm \sigma}\, \theta ({\bm v}_{12} \cdot \widehat{\bm \sigma}) {\bm v}_{12} \cdot \widehat{\bm \sigma} \left[ \delta ({\bm r}_{12} - {\bm \sigma}) \alpha^{-2} b_{\bm \sigma}^{-1} (1,2) - \delta ({\bm r}_{12} + {\bm \sigma}) \right].
\end{equation}

In this expression, $\theta(x)$ is the Heaviside step function, ${\bm v}_{12} \equiv {\bm v}_{1}-{\bm v}_{2}$ is the relative velocity of the two particles, ${\bm r}_{12} \equiv {\bm r}_{1}-{\bm r}_{2}$ their relative position vector, $d \widehat{\bm \sigma}$ is the solid angle element for the unit vector $\widehat{\bm \sigma}$, pointing from the center of particle $2$ to the center of particle $1$ at contact, and ${\bm \sigma}= \sigma \widehat{\bm \sigma}$. Finally, $b_{\bm \sigma}^{-1}(1,2)$ is an operator changing all the velocities ${\bm v}_{1}$ and ${\bm v}_{2}$ to its right into the pre-collisional values ${\bm v}^{*}_{1}$ and ${\bm v}^{*}_{2}$ given by
\begin{eqnarray}
\label{2.4}
{\bm v}^{*}_{1} & = & {\bm v}_{1} - \frac{1+ \alpha}{2 \alpha}\, {\bm v}_{12} \cdot \widehat{\bm \sigma} \widehat{\bm \sigma},
 \nonumber \\
{\bm v}^{*}_{2} & = & {\bm v}_{2} + \frac{1+ \alpha}{2 \alpha}\, {\bm v}_{12} \cdot \widehat{\bm \sigma} \widehat{\bm \sigma}.
\end{eqnarray}
From the one-particle distribution function, the macroscopic number of particles density, $n({\bm r},t)$, flow velocity, ${\bm u} ({\bm r},t)$, and granular temperature, $T({\bm r},t)$, are defined as
\begin{equation}
\label{2.5}
 n({\bm r},t) = \int d{\bm v}\, f({\bm r},{\bm v},t),
\end{equation}
\begin{equation}
\label{2.6}
n({\bm r},t) {\bm u}({\bm r},t) = \int d{\bm v}\, {\bm v} f({\bm r},{\bm v},t),
\end{equation}
\begin{equation}
\label{2.7}
\frac{d}{2} n({\bm r},t) T({\bm r},t) = \int d{\bm v}\, \frac{m}{2} \left[ {\bm v} -{\bm u}({\bm r},t) \right]^{2} f({\bm r},{\bm v},t).
\end{equation}
Because of the energy dissipation in collisions, there is no equilibrium state for the granular gas. Instead, there is a special state of homogeneous cooling (HCS), for which the system is translationally invariant and the temperature decreases monotonically in time. In the kinetic theory description of the HCS, it is assumed that  all the time dependence of the distribution function occurs through the granular temperature, so that it has the form \cite{GyS95}
\begin{equation}
\label{2.8}
f_{H}({\bm v},t) = n v_{0}^{-d}(t) \phi (c),
\end{equation}
where
\begin{equation}
\label{2.9}
v_{0}(t) \equiv \left( \frac{2T(t)}{m} \right)^{1/2}
\end{equation}
is a thermal velocity, and $\phi(c)$ is an isotropic function of the scaled velocity
\begin{equation}
\label{2.10}
{\bm c} \equiv \frac{\bm v}{v_{0}(t)}\, .
\end{equation}
From Eqs. (\ref{2.1}) and (\ref{2.8}) it is easily obtained
\begin{equation}
\label{2.11}
\frac{\partial T(t)}{\partial t} = - \zeta (t) T(t),
\end{equation}
with the cooling rate $\zeta(t)$ being given by
\begin{equation}
\label{2.12}
\zeta(t) = \frac{ \sigma^{d-1} \pi^{(d-1)/2} (1-\alpha^{2})n g_{e}(n)T(t)^{1/2}}{ (2m)^{1/2} \Gamma \left( \frac{d+3}{2} \right) d} \int d{\bm c}_{1} \int d{\bm c}_{2}\, c_{12}^{3} \phi (c_{1}) \phi(c_{2}).
\end{equation}
In the above equation, $g_{e}(n)$ is the equilibrium pair correlation function of two particles at contact, evaluated at the uniform local density $n$. The distribution function $f_{H}({\bm v},t)$ obeys the equation
\begin{equation}
\label{2.13}
\frac{\partial f_{H}(t)}{\partial t}= - \zeta(t) T(t) \frac{ \partial f_{H}(t)}{\partial T} = g_{e}(n) J_{B} [{\bm v}|f_{H},f_{H}],
\end{equation}
where $J_{B}$ is the inelastic Boltzmann collision operator defined by
\begin{equation}
\label{2.14}
J_{B}[f_{1},f_{2}] \equiv \sigma^{d-1} \int d{\bm v}_{2} \int d\widehat{\bm \sigma}\, \theta ({\bm v}_{12} \cdot \widehat{\bm \sigma}) {\bm v}_{12} \cdot \widehat{\bm \sigma} \left[b_{\bm \sigma}^{-1}(1,2) -1 \right]f_{1}({\bm v}_{1}) f_{2}({\bm v}_{2}).
\end{equation}

The explicit form of $\phi(c)$ is only known approximately. In the first Sonine approximation it reads \cite{GyS95,vNyE98},
\begin{equation}
\label{2.15}
\phi(c)= \pi^{-d/2} e^{-c^{2}} \left[ 1 + a_{2} S^{(2)}(c^2) \right],
\end{equation}
with
\begin{equation}
\label{2.16}
S^{(2)}(c^{2}) = \frac{d(d+2)}{8}- \frac{d+2}{2}\, c^{2} + \frac{c^{4}}{2}
\end{equation}
and
\begin{equation}
\label{2.17}
a_{2}\approx \frac{16(1-\alpha)(1-2 \alpha^{2})}{9+24d+(8d-41)\alpha+30 \alpha^{2} -30 \alpha^{3}}\, .
\end{equation}
In the same approximation, the HCS cooling rate is given by
\begin{equation}
\label{2.18}
\zeta(t) \approx \frac{2 \pi^{(d-1)/2} (1-\alpha^{2})n g_{e}(n)\sigma^{d-1}}{\Gamma \left( d/2 \right)d} \left( 1+ \frac{3 a_{2}}{16} \right) \left( \frac{T(t)}{m} \right)^{1/2}.
\end{equation}

\section{The steady self-diffusion state}
\label{s3}
Consider now that some of the particles are labeled, and the system as a whole is in the HCS. The Enskog equation for the time evolution of the one-particle distribution function of labeled particles, $f_{l}({\bm r}, {\bm v},t)$, is
\begin{equation}
\label{3.1}
\left( \frac{\partial}{\partial t} +{\bm v} \cdot \frac{\partial}{\partial {\bm r}} \right) f_{l}({\bm r}, {\bm v},t) = g_{e}(n) J_{B}[{\bm r},{\bm v}|f_{l},f_{H}].
\end{equation}
It is worth to stress that $f_{H}$ is the distribution function for all particles in the system, i.e. including both labeled and non-labeled particles. Also, the pair correlation function at contact is evaluated at the total local density of particles. This is because each of the particles feels in its motion a global bath of particles in the HCS, independently of the particle being labeled or non-labeled. Therefore, Eq.\ (\ref{3.1}) does not imply any additional hypothesis or approximation other that those required for the accuracy of Eq. (\ref{2.1}). The one-particle distribution function of non-labeled particles, $f_{nl}({\bm r},{\bm v},t)$, is
\begin{equation}
\label{3.2}
f_{nl}({\bm r},{\bm v},t)= f_{H}({\bm v},t)-f_{l}({\bm r},{\bm v},t).
\end{equation}
Then Eqs. (\ref{3.1}) and (\ref{2.13}) yield
\begin{equation}
\label{3.3}
\left( \frac{\partial}{\partial t} +{\bm v} \cdot \frac{\partial}{\partial {\bm r}} \right) f_{nl}({\bm r}, {\bm v},t) = g_{e}(n) J_{B}[{\bm r},{\bm v}|f_{nl},f_{H}],
\end{equation}
showing the required symmetry of the theory with regards to labeled and non-labeled particles. The number density of labeled particles, $n_{l}({\bm r},t)$, is
\begin{equation}
\label{3.4}
n_{l}({\bm r},t)= \int d{\bm v}\, f_{l}({\bm r},{\bm v},t),
\end{equation}
and integration of Eq.\ (\ref{3.1}) leads to
\begin{equation}
\label{3.5}
\frac{\partial n_{l}}{\partial t}+ \frac{\partial}{\partial {\bm r}} \cdot {\bm j}_{l}({\bm r},t)=0,
\end{equation}
where ${\bm j}_{l}({\bm r},t)$ is the flux of labeled particles defined as
\begin{equation}
\label{3.6}
{\bm j}_{l}({\bm r},t) = \int d{\bm v}\ {\bm v} f_{l}({\bm r},{\bm v},t) = n_{l}({\bm r},t) {\bm u}_{l}({\bm r},t).
\end{equation}
The last equality defines the local average velocity of labeled particles, ${\bm u}_{l}({\bm r},t)$. In the following, attention will be restricted to a state characterized by an homogeneous flux of labeled particles in a given direction, arbitrarily taken as the $z$ axis. Equation (\ref{3.5}) implies that the density profile of labeled particles is stationary, $n_{l}=n_{l}(z)$. Assuming that there are no gradients perpendicular to the $z$ axis, Eq. (\ref{3.1}) reduces to
\begin{equation}
\label{3.7}
\frac{\partial f_{l}}{\partial t}+v_{z} \frac{\partial f_{l}}{\partial z}= g_{e}(n) J_{B}[z,{\bm v}|f_{l},f_{H}].
\end{equation}
Prompted by the results obtained for elastic, molecular systems \cite{ByR13a}, we search for a solution of the above equation having the form
\begin{equation}
\label{3.8}
f_{l}(z,{\bm v},t)=\left[ n_{l}(z)+ \chi ({\bm c}) \right] \frac{ f_{H}({\bm v},t)}{n},
\end{equation}
where $\chi ({\bm c})$ is a function of the scaled velocity ${\bm c}$ defined in Eq.\ (\ref{2.10}). It is assumed that this function does not depend on the density of labeled particles $n_{l}$ and, even more, that it is position independent. Of course, the consistency of this assumption will come from the existence of a solution of the form assumed above. Note that the distribution given by Eq. (\ref{3.8}) belongs to the class of distribution functions called {\em normal} \cite{RydL77}, in which all the position and time dependence occurs through the hydrodynamic fields, in this case the density of labeled particles and the temperature of the system. Consistency of Eqs.\ (\ref{3.4}) and (\ref{3.8}) requires that
\begin{equation}
\label{3.9}
\int d{\bm c}\, \chi({\bm c}) \phi (c)=0.
\end{equation}
Substitution of Eq.\ (\ref{3.8}) into Eq.\ (\ref{3.7}), taking into account Eq.\ (\ref{2.13}), yields
\begin{equation}
\label{3.10}
\frac{\partial}{\partial t} \left[ \chi ({\bm c}) f_{H}({\bm v},t) \right] + v_{z} f_{H} \frac{\partial n_{l}}{\partial z}= g_{e} (n) J_{B} [{\bm v}| \chi f_{H},f_{H}].
\end{equation}
Since all the position dependence in the above equation is through $n_{l}(z)$, it follows that the only density profile compatible with the assumed form of the distribution function of labeled particles as given by Eq.\ (\ref{3.8}) is a linear one, namely
\begin{equation}
\label{3.11}
n_{l}=az+b,
\end{equation}
$a$ and $b$ being arbitrary constants to be determined in each case from the boundary  conditions. Of course, because of the particular system being considered, it must be $0 \leq n_{l} \leq n$ everywhere. Then, Eq.  (\ref{3.10}) is equivalent to
\begin{equation}
\label{3.12}
\zeta (t) T(t) \frac{\partial}{\partial T(t)}\, \left[ \chi ({\bm c}) f_{H}({\bm v},t)\right] + g_{e}(n) J_{B}[{\bm v}|\chi f_{H},f_{H}]=v_{z} a f_{H}({\bm v},t).
\end{equation}
The solution $\chi({\bm c})$ of the above equation has to be proportional to $a$. Let us write
\begin{equation}
\label{3.13}
\chi ({\bm c}) f_{H} ({\bm v},t)= n a B({\bm v}).
\end{equation}
The function $B({\bm v})$ will also depend on time through $T(t)$, but this will not be indicated explicitly. Use of Eq.\ (\ref{3.13}) into Eq. (\ref{3.12}) leads to
\begin{equation}
\label{3.14}
\zeta(t) T(t) \frac{\partial B(t)}{\partial T(t)} + g_{e}(n) J_{B}[{\bm v}|B,f_{H}]= v_{z}\, \frac{f_{H}({\bm v},t)}{n}\, .
\end{equation}
Moreover, from Eqs.\ (\ref{3.6}) and (\ref{3.8}) it follows that
\begin{equation}
\label{3.15}
j_{l,z}= - D(t) a,
\end{equation}
with the self-diffusion coefficient $D$ identified as
\begin{equation}
\label{3.16}
D(t)=- \int d{\bm v}\, v_{z}B({\bm v}).
\end{equation}
Equation (\ref{3.15}) has the same form as the Fick law derived when studying the self-diffusion equation to Navier-Stokes order by means of the Chapmann-Enskog procedure,  and it has been solved in the first Sonine approximation \cite{BRCyG00}. In the Appendix, an alternative approximation is discussed. It has proven to lead to slightly more accurate expressions for the transport coefficients of a granular gas, especially for strong inelasticity \cite{BMyG12}. The result for the self-diffusion coefficient is
\begin{equation}
\label{3.17}
\frac{D(t)}{D_{0}(t)}= \frac{4}{(1+\alpha)^2 \left[ 1+ \frac{3 a_{2}(\alpha)}{16} \right]}\, ,
\end{equation}
where $D_{0}(t)$ is the elastic value of the self-diffusion coefficient for hard spheres or disks evaluated at the temperature $T(t)$,
\begin{equation}
\label{3.18}
D_{0}(t) = \frac{\Gamma \left(d/2 \right)d}{4 \pi^{(d-1)/2} n g_{e}(n)\sigma^{d-1}}\, \left ( \frac{T(t)}{m} \right)^{1/2},
\end{equation}
and $a_{2}(\alpha)$ is given by Eq.\ (\ref{2.17}). In the same approximation, the expression of the one-particle distribution function of labeled particles is
\begin{equation}
\label{3.19}
f_{l}(z,{\bm v},t)= \left[ n_{l} - \frac{2 a D(t)}{v_{0}(t)}\, c_{z} \right] v_{0}^{-d}(t) \phi(c).
\end{equation}
This expression deserves some comments. It is seen that for large positive values of $c_{z}$ it becomes negative, what is unphysical. This should not be understood as a signal of a limitation of Eq.\ (\ref{3.14}) or even as an indication of the absence of a solution of the inelastic Enskog equation of the form introduced in Eq.\ (\ref{3.8}). Actually, Eq.\ (\ref{3.8}) verifies the solubility conditions since it formally agrees with one of the equations derived by the Chapman-Enskog procedure \cite{BRCyG00}. The anomalous behaviour of Eq.\ (\ref{3.19}) for large values of $c_{z}$ is a consequence of the approximation made to derive it, namely Eq.\ (\ref{a.12}).

Next consider the velocity moments of the distribution of labeled particles $\mu_{k} (z,t)$ defined as
\begin{equation}
\label{3.20}
\mu_{k}(z,t) \equiv \int d{\bm v}\, v_{z}^{k} f_{l} (z,{\bm v},t).
\end{equation}
In particular, it is
\begin{equation}
\label{3.21}
\mu_{0}(z)=n_{l}(z)= az+b, \quad \mu_{1}=j_{l,z}=-Da, \quad \mu_{2}(t)=\frac{n_{l} T(t)}{m}\, .
\end{equation}
In general, use of  Eq. (\ref{3.19}) leads to
\begin{equation}
\label{3.22}
\mu_{k}= v_{0}^{k}(t) \left[ n_{l}(z) \nu_{k} - \frac{2D(t)a}{v_{0}(t)}\, \nu_{k+1} \right],
\end{equation}
with $\nu_{k}$ given by
\begin{equation}
\label{3.23}
\nu_{k} \equiv \int d{\bm c}\, c_{z}^{k} \phi(c).
\end{equation}
The isotropy of $\phi(z)$ implies that $\nu_{k}=0$ for odd $k$. Moreover, Eq. (\ref{2.8}) implies that $\nu_{0}=1$ and $\nu_{2}=1/2$. The moments $\nu_{4}$ and $\nu_{6}$ are related to the coefficients appearing in the Sonine expansion of $\phi (c)$. In particular, $\nu_{4}$ is determined by the coefficient $a_{2}$ appearing in Eq.\ (\ref{2.15}) and given in Eq.\ (\ref{2.17}) \cite{GyS95,vNyE98}. The known expressions  for both $\nu_{4}$ and $\nu_{6}$ are in good agreement with simulation results, at least for not very strong inelasticity \cite{BRyC96,ByP06}.

\section{Simulation results}
\label{s4}
To check the accuracy of the predictions derived above, numerical solutions of the inelastic Boltzmann equation have been generated by using the direct simulation Monte Carlo method (DSMC) \cite{Bi94}. This is a particle simulation algorithm designed to mimic the dynamics of the particles in a low density gas, and it has proven to be extremely efficient and accurate. Since the method has been reviewed many times in the literature \cite{Ga00}, it will not be described here.

The aim of the simulations was to investigate the existence of a solution of the kinetic equation describing the uniform self-diffusion state, and the accuracy of its predicted properties. On the other hand, the focus was not on delimiting  the validity of the inelastic Enskog equation. For this reason, and to decrease  the statistical uncertainty of the simulation results, the DSMC method was used. Note that the difference between the Boltzmann and the Enskog description for this particular state is the constant factor of the equilibrium pair correlation of the fluid at (homogeneous) equilibrium, $g_{e}(n)$, in front of the collision operator. As a consequence, proving the existence of a solution of the Enskog equation for this state is equivalent to establishing it for the Boltzmann equation.

In the simulations, a system of hard spheres confined between two plates located at $z=0$ and $z=L$, respectively, was considered. To generate a non-homogeneous density of labeled particles along the $z$-direction, nonlocal boundary conditions were used. They are conceptually similar to those introduced many years ago by Lees and Edwards \cite{LyE72} to generate the uniform shear flow state, and with slight differences they are the same as those employed by Erpenbeck and Wood \cite{EyW77}.

The boundary conditions are as follows. When a particle reaches the wall at $z=L$, it is reinjected in the system at $z=0$ with the same velocity, but it will be labeled with probability $p$, independently of whether it was or was not labeled before. Also, a particle leaving the system through the the plate at $z=0$ is reinserted through $z=L$ with a probability $q$ of being labeled. In the simulations to be reported in the following, the choice $p+q=1$ was made, in order to make the roles of labeled and unlabeled particles symmetric. It is clear that the effect of this label assignment at the boundaries of the system is to generate a gradient in the density of labeled particles. The particular choice $p=1$ corresponds to a reservoir of labeled particles at $z=0$ and one of unlabeled particles at $z=L$. In the long time limit, it is expected that the density of labeled particles will reach a steady profile. On the other hand, the shape of the profile is by no means imposed by the boundary conditions, but determined by the transport equations of the system. To modify the gradient of the density of labeled particles, both the value of $p$ and the size $L$ of the system were varied in the simulations.

The study by simulation techniques of self-diffusion in a granular gas in the HCS presents two additional difficulties, as compared with the same study in a molecular, elastic system at equilibrium. First, the HCS becomes unstable when the linear size of the system is larger than a critical value that depends on the coefficient of normal restitution $\alpha$ \cite{GyZ93,BRyM98}. Second, the system is continuously cooling, and after some time the typical velocities of the particles become very small and, consequently, the statistical errors very large. This latter problem can be solved by mapping the properties of the HCS onto the dynamics around a steady state, by means of a change in the time scale \cite{Lu01,BRyM04}. This steady representation of the HCS has been employed in all the simulations reported in the following. The instability of the HCS implies that systems smaller than the critical size have to be used. For the interval of inelasticity considered here, $0.65 \leq \alpha \leq 0.98$, the critical size $L_{c}$ lies in the interval $15.39 \lambda \leq L_{c} \leq 54.43 \lambda$, where $\lambda = ( \sqrt{2}  \pi n \sigma^{2})^{-1}$ is the mean free path.

The simulations started  in all  cases with the same number of labeled and unlabeled particles, being both uniformly distributed in the system. The initial velocity distribution was Gaussian with zero mean and granular temperature set equal to unity. After some transient time period, the system always reached a steady state. The results to be presented have been averaged over 200 trajectories and also during a period of time of typically 2000 collisions per particle. To compute the different properties of the system, it was divided into layers perpendicular to the $z$ axis of width $\Delta z= \lambda$.

\begin{figure}
\begin{tabular}{cc}

\includegraphics[scale=0.35,angle=0]{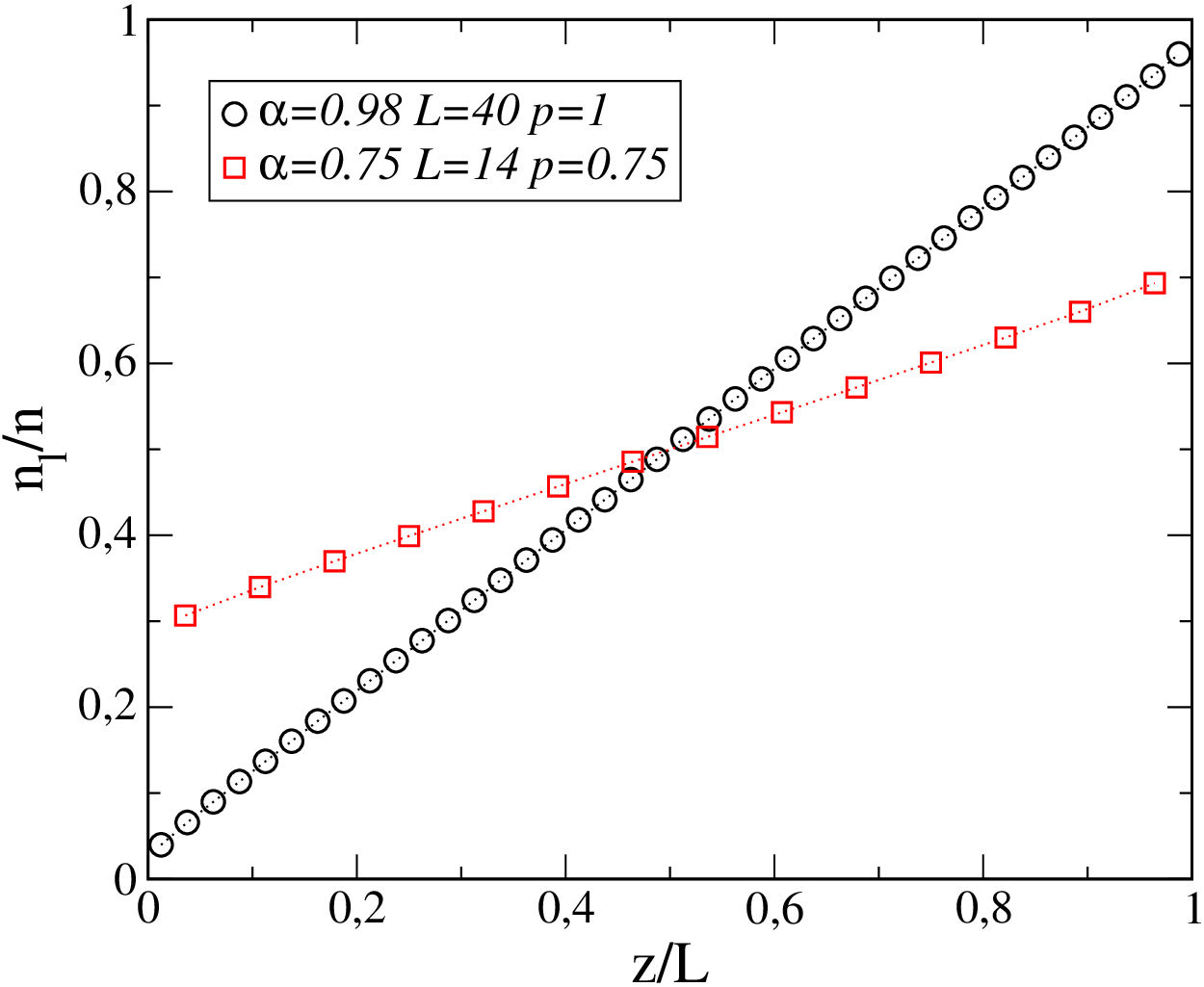} & \quad \quad
\includegraphics[scale=0.35,angle=0]{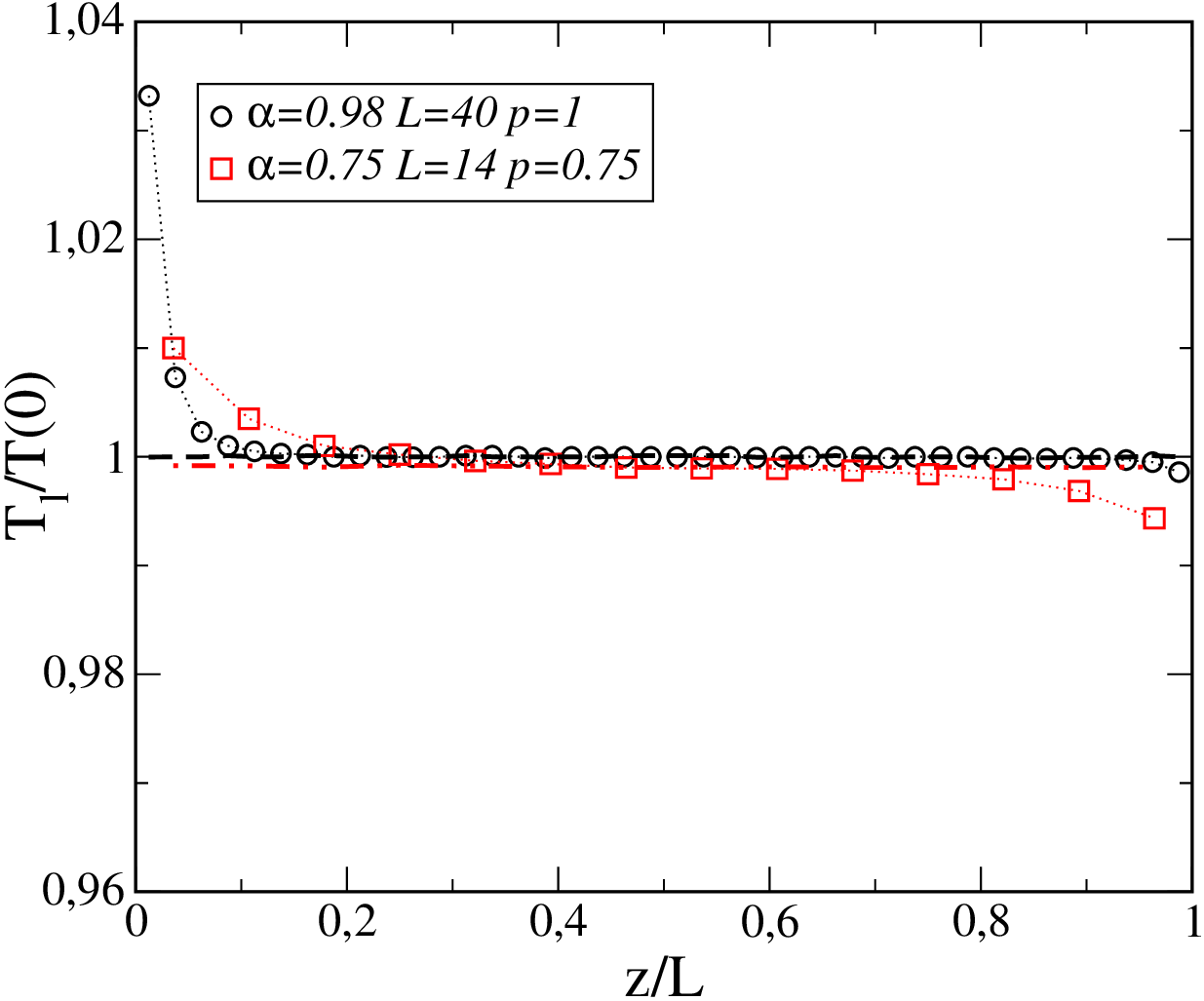}
\end{tabular}
\caption{Relative  number density $n_{l}/n$ (left) and temperature $T_{l}/T(0)$ (right) profiles of labeled hard spheres in two steady states corresponding to the set of parameters indicated in the inset. The density is scaled with the total number density density $n$ and the temperature is normalized with the initial temperature of the system. In the right plot, the total temperature profiles in both states are also shown (dashed and dot-dashed lines). \label{fig1}}
\end{figure}

As an example, the steady density and temperature profiles for the labeled particles are shown in Fig. \ref{fig1} for two sets of values of the parameters, namely $\alpha=0.98$, $L = 40 \lambda$ and $p=1$ in one case, and $\alpha=0.75$, $L= 14 \lambda$, and $p=0.75$, in the other. In both systems, the density profile of labeled particles is clearly linear, as predicted by the theory. Also the temperature of the labeled particles is uniform and agrees with the global temperature of the system, aside from a boundary layer. The small deviation of the total steady temperature from the initial value $T(0)$ is due to the small error of the expression for the cooling rate in the first Sonine approximation, an expression that is used to construct the steady representation of the HCS \cite{BRyM04}. In any case, this is a technical detail without physical implications. Similar density and temperature profiles were obtained in all the simulations being reported.

From the fit to a straight line of the density profiles, the value of the uniform density gradient $a$ can be computed in each case. This value does not agree exactly with $(p-q)/L=(2p-1)/L$, due to the existence of boundary layers next to the walls. Also the flux of particles has been measured in the simulations, and verified that it is actually uniform. Then, the simulation values of the self-diffusion coefficient $D$ have been obtained in each case by means of  Eq.\ (\ref{3.15}). No dependence of the self-diffusion coefficient on the density gradient $a$ was observed, in agreement with the theoretical prediction. The measured values of the density gradient were roughly in the range $ 0.011 \leq a \lambda /n \leq 0.067$.

\begin{figure}
\includegraphics[scale=0.4,angle=0]{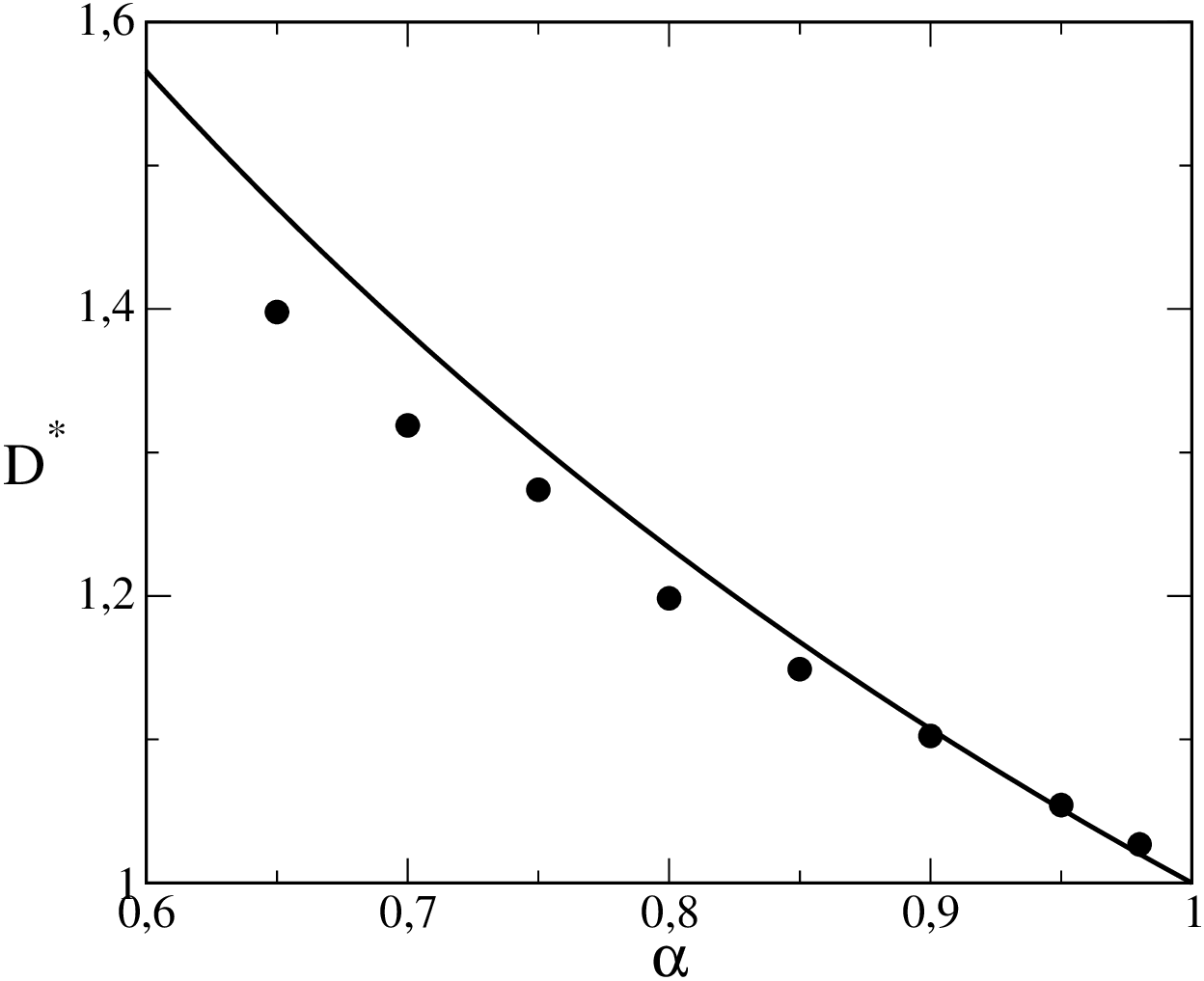}
\caption{Dimensionless reduced self-diffusion coefficient $D^{*}$  of a system in the uniform self-diffusion state, as a function of the coefficient of normal restitution $\alpha$. The symbols are the simulation results, while the solid line is the theoretical prediction given in the main text. \label{fig3}}
\end{figure}

In Fig.\ \ref{fig3} the reduced coefficient of self-diffusion, $D^{*} \equiv D(t) / D_{0}(t) $, is plotted as a function of the coefficient of normal restitution $\alpha$. The symbols are the simulation results, while the solid line is the theoretical prediction given by Eq.\ (\ref{3.17}). It must be emphasized that the correction due to the deviation from a Gaussian of the  distribution function of the HCS, given by the term proportional to $a_{2}$, is  imperceptible on the scale of the figure.  Also the difference between Eq.\ (\ref{3.17}) and the Chapman-Enskog result to Navier-Stokes order \cite{BRCyG00} is negligible and, therefore, it is hard to say which one fits more accurately the simulation values. The results for the self-diffusion coefficient obtained here are similar to those obtained by other methods to Navier-Stokes order \cite{BRCyG00,LByD02}. The inelasticity of collisions enhances diffusion, and this is not a minor effect. The systematic increase of the discrepancy between the
simulation results and the theoretical prediction as the inelasticity of the collisions increases, is due to the loss of accuracy of the approximation introduced in the Appendix to evaluate the self-diffusion coefficient.

The local velocity distribution of labeled particles has also been investigated in the simulations. First, the profiles of the velocity moments $\mu_{k}$ defined in Eq.\, (\ref{3.20}) have been considered. The theoretical prediction for them is given by Eq.\ (\ref{3.22}) and involves the velocity moments of the HCS, $\nu_{k}$, defined in Eq.\ (\ref{3.23}). Although, as already mentioned, $\nu_{0}=1$ and $\nu_{2}= 1/2$, and accurate expressions are known for the moments $\nu_{4}$ and $\nu_{6}$, here values obtained from the simulations themselves will be used, in order to avoid additional sources of discrepancy, focussing on the comparison of the predictions associated to the uniform self-diffusion state. Since the moments $\nu_{4}$ and $\nu_{6}$ have been extensively studied elsewhere \cite{BRyC96,ByP06}, the results obtained for them in the simulations will not be reproduced here.

\begin{figure}
\begin{tabular}{cc}
\includegraphics[scale=0.35,angle=0]{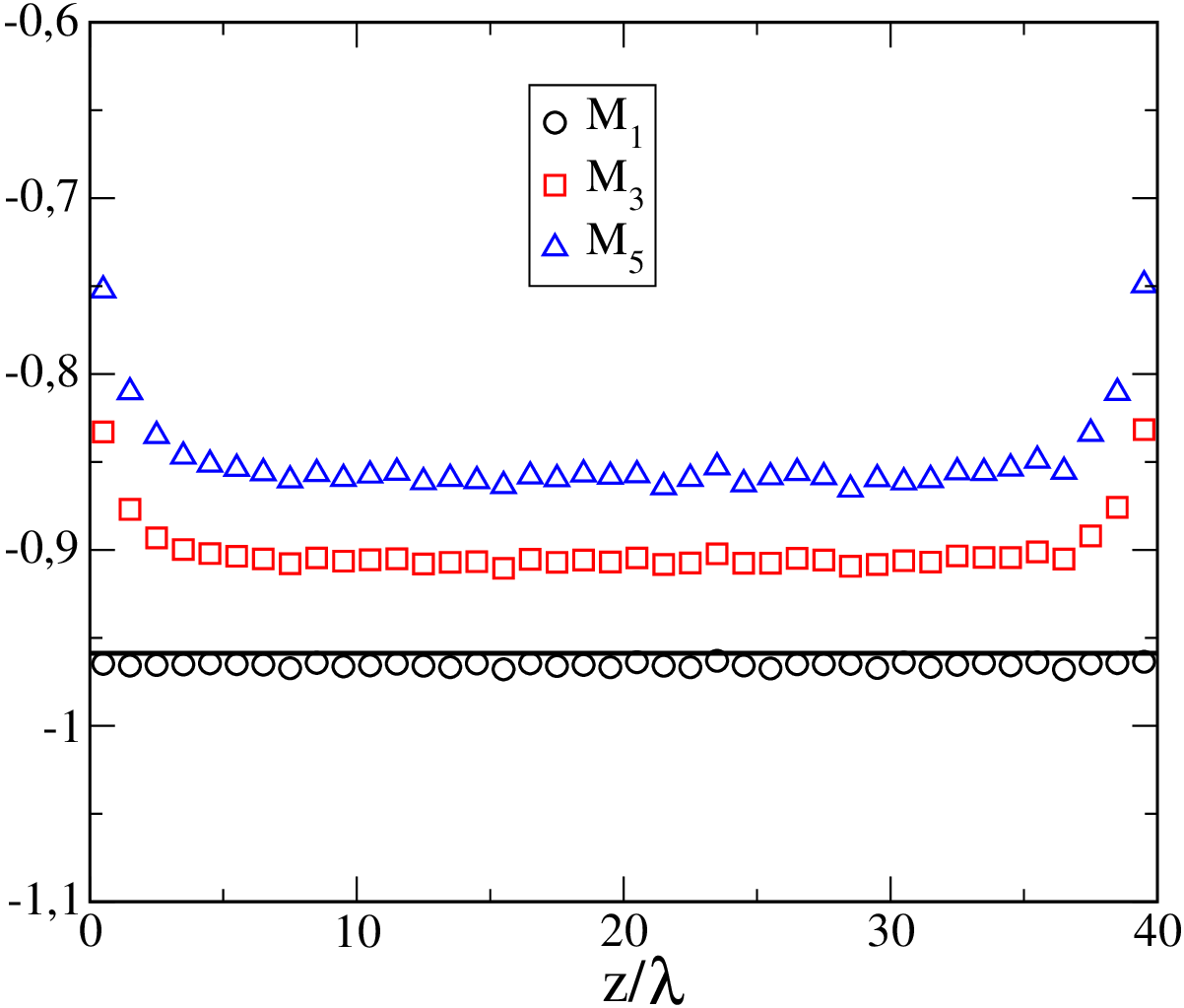} & \quad \quad
\includegraphics[scale=0.35,angle=0]{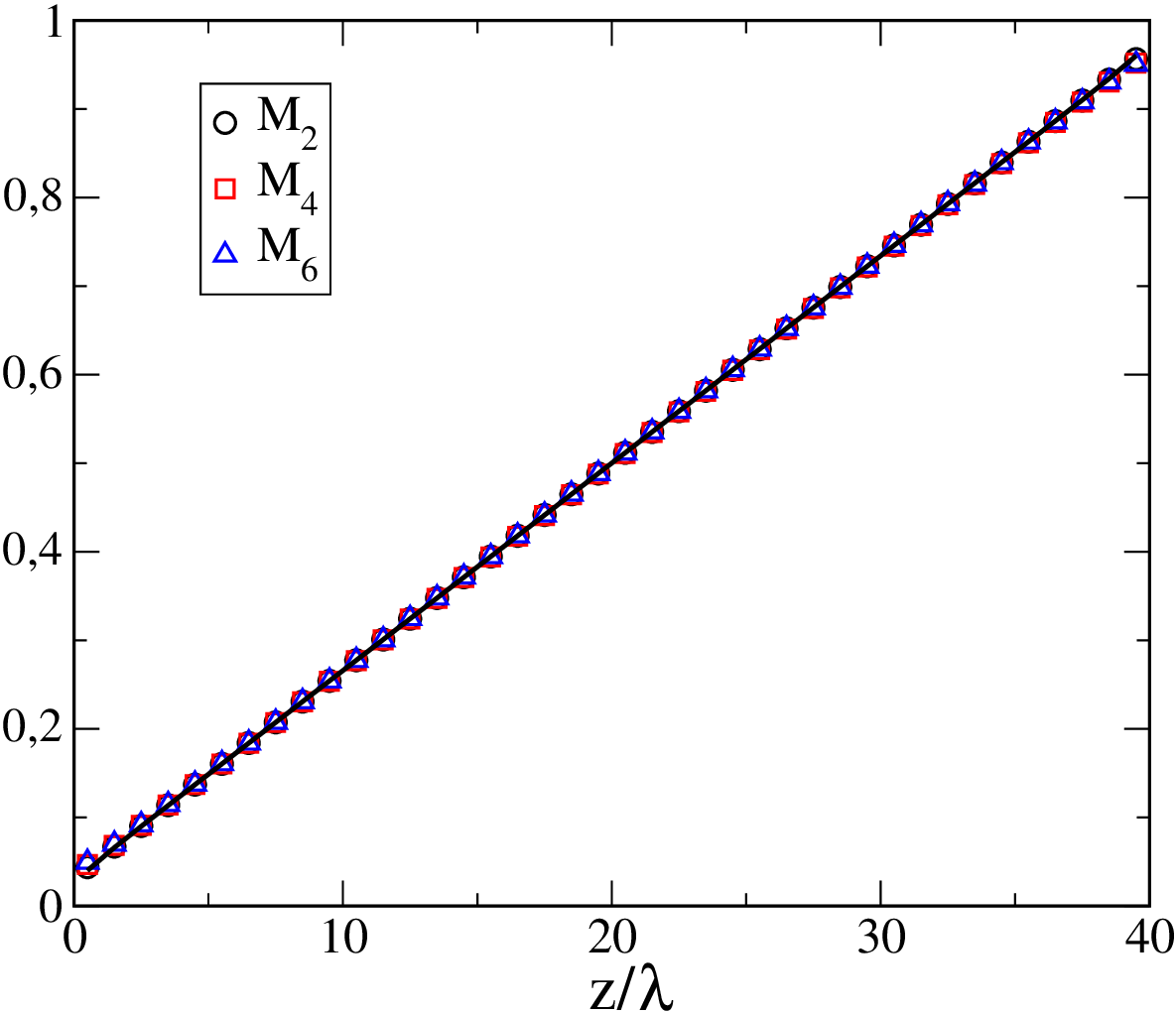}
\end{tabular}
\caption{Dimensionless scaled odd (left) and even (right) velocity moments of the distribution of labeled particles for a system with $\alpha=0.98$, $p=1$, and $L=40 \lambda$. In each plot, the solid line is the theoretical prediction, which is $- 2 D(t)/v_{0}(t)\lambda $ for the even moments, and $n_{l}(z)/n$ for the even ones. \label{fig4}}
\end{figure}

Once the HCS values of the moments  are known, scaled  moments for the uniform self-diffusion state are defined as
\begin{equation}
\label{4.1}
M_{2k+1}= \frac{\mu_{2k+1}}{a v_{0}^{2k+1}(t) \nu_{2k+2} \lambda}\, ,
\end{equation}
\begin{equation}
\label{4.2}
M_{2k}= \frac{\mu_{2k}}{n v_{0}^{2k}(t) \nu_{2k}}\, .
\end{equation}
With this scaling, the theoretical prediction, Eq.\ (\ref{3.22}), is that the odd scaled moments
$M_{2k+1}$ are constant and equal to $-2 D(t)/v_{0}(t) \lambda$, while the even scaled moments $M_{2k}$ are equal to the relative density of labeled particles $n_{l}(z)/n$, for all $k$. In Fig.\ \ref{fig4}, the first scaled moments are shown for a system with $\alpha=0.98$, $L/\lambda=40$, and $p=1$. The solid lines are the theoretical predictions in each case. It is seen that the agreement between theory and simulations is very good for all the even moments, while a systematic discrepancy is observed for the odd moments, although it is true that they are uniform in the bulk of the system, i.e. outside the boundary layers, as predicted. Moreover, the discrepancy is larger the larger $k$. Probably, it is also due to the approximation introduced in the Appendix, that not only affects the value of the self-diffusion coefficient, but also the functional dependence of the distribution function on the velocity.

\begin{figure}
\begin{tabular}{cc}
\includegraphics[scale=0.35,angle=0]{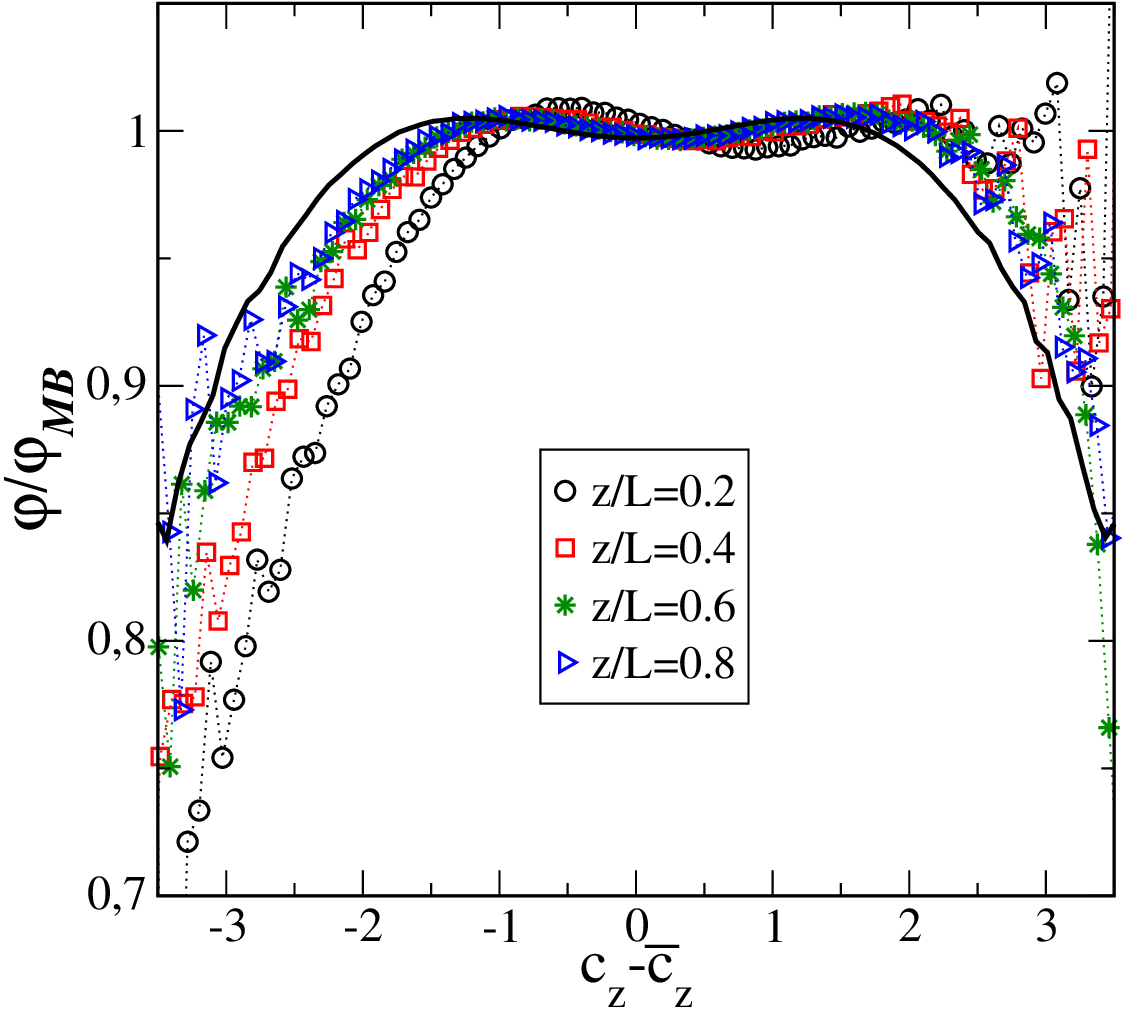} & \quad \quad
\includegraphics[scale=0.35,angle=0]{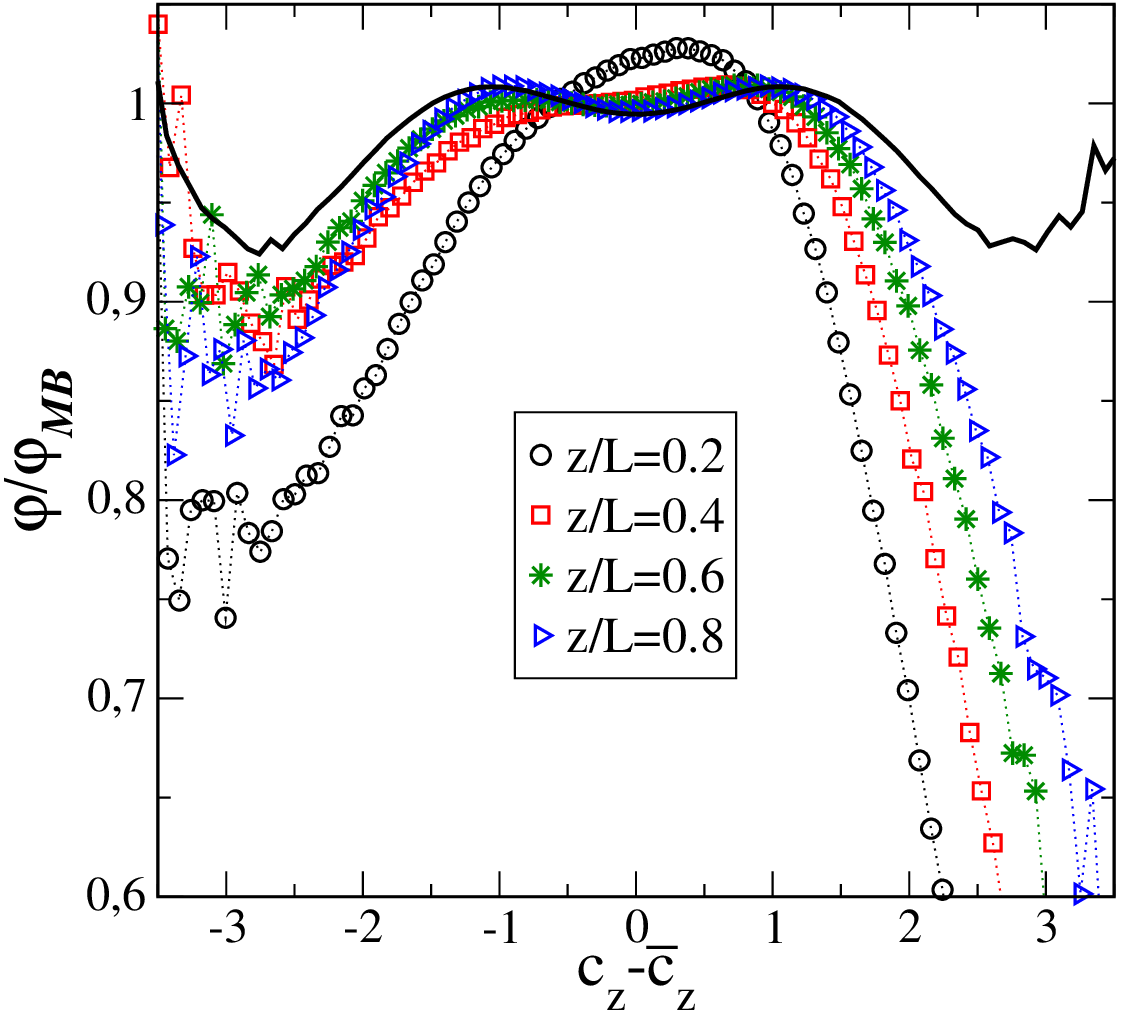}
\end{tabular}
\caption{Probability distribution of the dimensionless scaled $z$ component of the velocity of labeled particles, $\varphi$, scaled with the Gaussian $\varphi_{MB}$ in the uniform self-diffusion state. The different symbols correspond to different layers centered at the positions indicated in the inset, and the solid line is the velocity distribution of the whole system, which is  in the HCS. The values of the parameters are $\alpha=0.98$, $L=40 \lambda$, and $p=1$ in the left plot, and $\alpha=0.75$, $L=14 \lambda$, and $p=1$ in the right one. \label{fig6}}
\end{figure}

Finally, the local velocity distribution of labeled particles  has been studied. For that purpose, four layers of width $\lambda$ located at $z=0.2L$, $z=0.4L$, $z=0.6L$, and $z=0.8 L$, respectively, have been considered, and the marginal distribution of the $z$ component of the velocity of labeled particles has been measured in each layer. It must be kept in mind that both the density and the average velocity of labeled particles depend on $z$ in the uniform self-diffusion state. The local velocity distributions of the $z$ component, $\varphi(c_{z})$, normalized to unity, are plotted in Fig. \ref{fig6}, for two different states, the second one exhibiting a much larger density gradient of labeled particles than the former. The velocities have been scaled with the thermal velocity $v_{0}(t)$, as in Eq. (\ref{2.10}) and the distributions have been divided by the Gaussian,
\begin{equation}
\label{4.3}
\varphi_{MB}(c_{z})= \frac{e^{-c_{z}^{2}}}{\sqrt{\pi}}\, .
\end{equation}

The different symbols in the figures correspond to the different positions of the layers in which the marginal velocity distribution has been measured in the simulations, as indicated in the inset. The solid lines are the distributions of the HCS in each case, also obtained from the simulations, using all the particles in the system. To render the comparison easier, the velocity origin has been displaced in each case with the average velocity, $\overline{c}_{z}(z)$. It is observed that the dependence on position of the velocity distribution function is not only through the local value of the mean velocity, but there is a clear change in the shape of the distribution. Moreover, the larger the labeled density gradient, the stronger the position dependence of the local velocity distribution.

The theoretical prediction for the one-particle distribution of labeled particles is given in Eq.\ (\ref{3.19}). From it, the marginal distribution for the $z$ component, $f_{l,z}(z,v_{z},t)$, is easily obtained by integration over $v_{x}$ and $v_{y}$. Define the function
\begin{equation}
\label{4.4}
\Phi (z,c_{z}) = \frac{f_{l,z}(z,v_{z},t)}{v_{0}^{-1}(t) \phi_{z}(c_{z})}\, - n_{l}(z).
\end{equation}
Here $\phi_{z}(c_{z})$ is the marginal velocity distribution for $c_{z}$ in the HCS. The theoretical prediction for $\Phi$ is
\begin{equation}
\label{4.5}
\Phi= - \frac{2 a D(t)}{v_{0}(t)}\, c_{z},
\end{equation}
which is independent from position and time.

\begin{figure}
\begin{tabular}{cc}
\includegraphics[scale=0.35,angle=0]{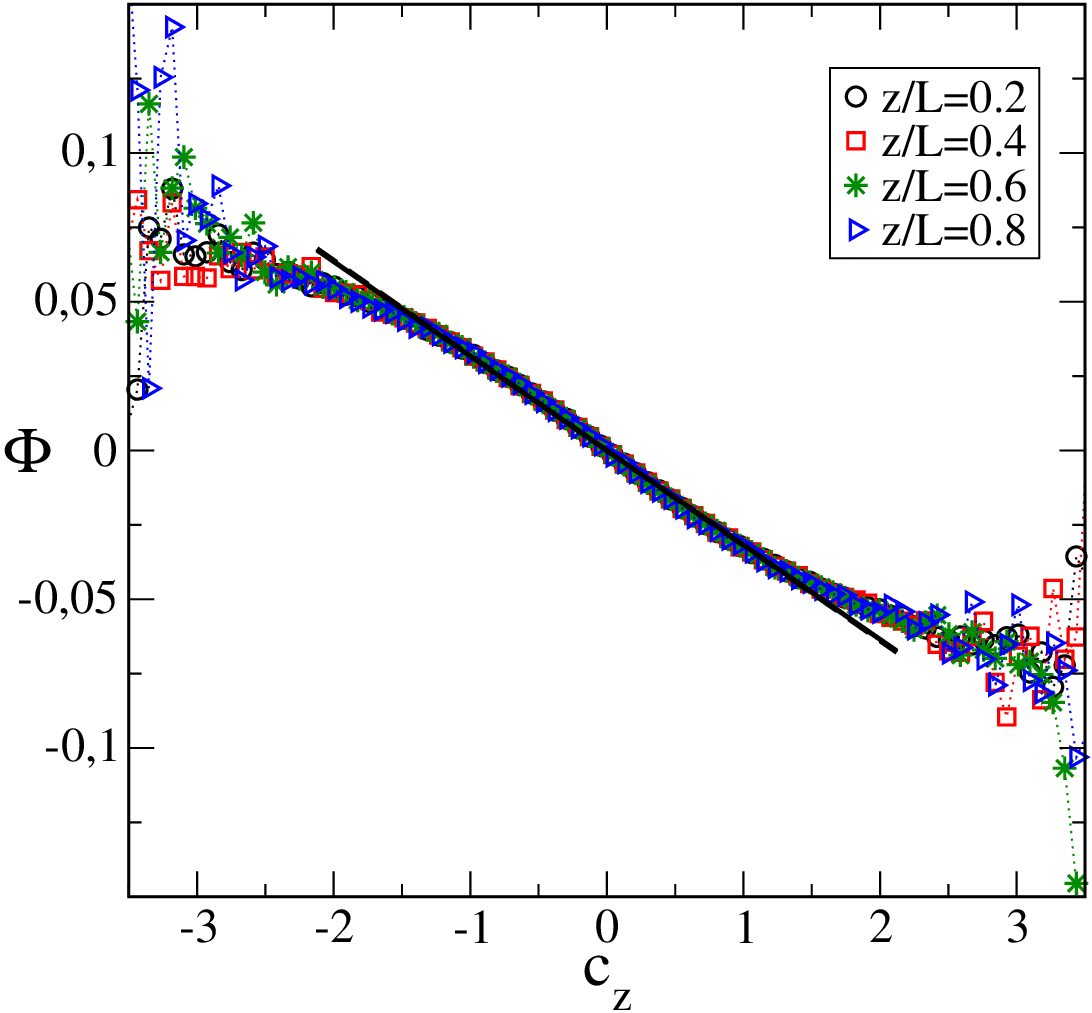} & \quad \quad
\includegraphics[scale=0.35,angle=0]{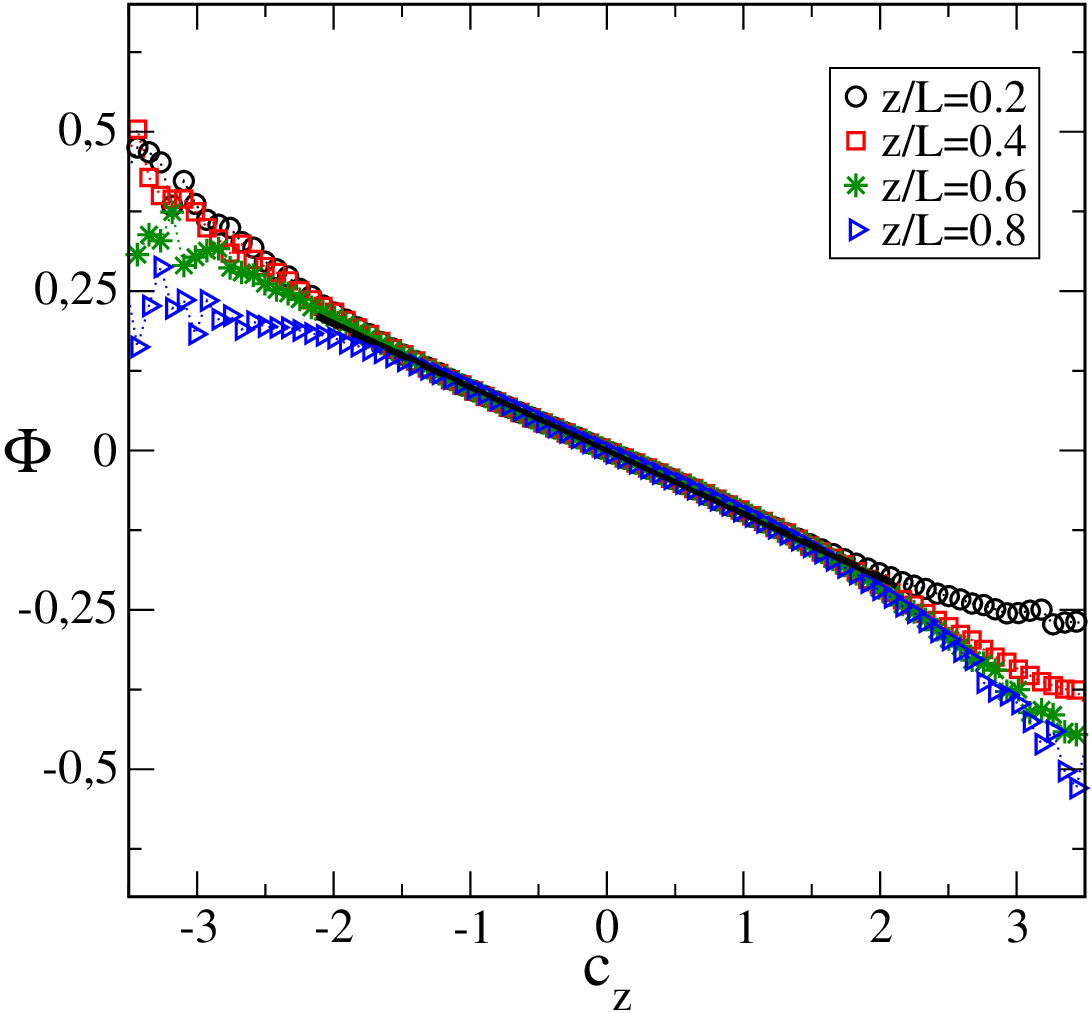}
\end{tabular}
\caption{Dimensionless scaled marginal distribution of labeled particles $\Phi$ for the same systems as in Fig.\ \protect{\ref{fig6}}. The solid lines are the theoretical prediction given by Eq.\ (\protect{\ref{4.5}}). \label{fig8}}
\end{figure}

The results for the function $\Phi (z,c_{z})$ obtained in the simulations for the same systems as in Fig.\ \ref{fig6}  are given in Fig. \ref{fig8}. When computing it, the function $\phi_{z}(c_{z})$ used has also been the one measured in the simulations. The different symbols correspond to layers of the system centered at different values of $z$, as indicated in the inset, while the solid lines are the theoretical predictions of Eq.\ (\ref{4.5}). With regards to the latter, there is no significant difference between using the measured value of the self-diffusion coefficient  or the theoretical prediction.  The curves obtained with both values are undistinguishable on the scale of the figure. A satisfactory agreement is seen in both cases. The observed discrepancies occurs for large velocities as compared with the typical thermal value $v_{0}$. For them, the approximation used to derive Eq.\ (\ref{3.19}) is not expected to be accurate. Also, the results show that the discrepancy between theory and simulations increases as the walls of the system are approached, due to boundary effects.

\section{Summary}
\label{s5}
In this paper, it has been shown that kinetic theory methods based on the Enskog equation for inelastic hard spheres or disks predict the existence of a self-diffusion state, in which the gas as a whole is in the HCS and the concentration of labeled particles is time-independent and exhibits a linear profile. Consistently, the flux of labeled particles is uniform. Moreover, this flux is proportional to the density gradient, independently of the magnitude of the latter, the inelasticity of collisions,  and the relative density of labeled particles. The associated transport coefficient is given by the same expression as obtained to Navier-Stokes order. The theoretical predictions have been compared with numerical simulations of the kinetic equation, and a good agreement has been found, even for the one-particle distribution function of labeled particles. To generate the state in the simulations, nonlocal boundary conditions, in the spirit of non-equilibrium simulation methods, have been used.

To put the above results in a proper context some comments seem appropriate:

1. The fact that the Boltzmann-Lorentz kinetic equation describing the evolution of the one-particle distribution of labeled particles be linear does not imply by itself that the flux of labeled particles has to be linear in their density gradient. There are many examples of linear kinetic equation leading to non-linear transport equations.

2. In this context, we do not claim that the self-diffusion process in the HCS of a granular gas of inelastic hard spheres or disks be always linear. In principle, the analysis here has been restricted to a particular self-diffusion state. The existence of this state is one of the results in this paper.

3. The above two comments also apply to the existence of linear higher-order terms, i.e. contributions to the particles flux involving higher-order gradients of the density of labeled particles, beyond Fick's law. The result here is that there is a special state in which the flux is proportional to the gradient of the density, independently of the concentration gradient.

4.  The linear density profile or, equivalently, the uniform flux of particles is not imposed by the simulation method, but generated by the system. Any shape of the density of labeled particles profile is, in principle, compatible with the imposed boundary conditions.

5. Actually, one is tempted to conclude that independently of the imposed boundary conditions, any state with a steady density profile of labeled particles in one direction, will exhibit a linear profile and the flux of particle will obey Fick's law.

6. The only way to test the presence or absence of rheological effects in the present state is by generating the density profile with the largest possible gradient. This has been done by taking equal to unity the probability of a particle being labeled (unlabeled) when reintroduced into the system through the boundary at $z=0$ ($z=L$). As discussed in the text, no rheological effects, linear or nonlinear, have been observed.

7. Finally, it is worth to stress that the test of the theory has not been restricted to the accuracy of the transport equation. The much more demanding comparison of the theoretical predictions and simulation results for the velocity moments and the velocity distribution itself has been carried out, and a good agreement has been found for the range of thermal velocities. The discrepancy for large velocities is probably due to the failure in that region of the approximation made to solve the kinetic equation.

An interesting question is whether the existence of the uniform self-diffusion state  and its properties derived here are a peculiarity of the (inelastic) Boltzmann-Enskog equation, or they also apply beyond the range of validity of the kinetic equation. Answering this question would require to use a Liouville description of the system and, to check the theory, perform molecular dynamics simulations.

\section{Acknowledgements}

This research was supported by the Ministerio de Educaci\'{o}n y Ciencia
(Spain) through Grant No. FIS2011-24460 (partially financed by FEDER funds).

\appendix*

\section{Evaluation of the self-diffusion coefficient and the distribution function}
In this Appendix, the approximation scheme leading to the expression  for the coefficient of self-diffusion $D(t)$ given in Eq. (\ref{3.17}) is described. It is convenient to introduce a dimensionless time scale $s$ by
\begin{equation}
\label{a.1}
s= \int_{0}^{t} dt^{\prime}\, \frac{v_{0}(t^{\prime})}{\ell}\, ,
\end{equation}
where $\ell \equiv (n \sigma^{d-1})^{-1}$ is proportional to the mean free path of the particles in the gas, and $v_{0}(t)$ is the thermal velocity defined in Eq.\ (\ref{2.9}). In the dimensionless units defined by the velocity ${\bm c}$ (see Eq.\ (\ref{2.10})) and $s$, Eq.\ (\ref{3.14}) reads
\begin{equation}
\label{a.2}
-\frac{\widetilde{\zeta}}{2} \frac{\partial}{\partial {\bm c}}\, \cdot \left[ {\bm c} \widetilde{B}({\bm c}) \right] + g_{e}(n) \widetilde{J}_{B} [{\bm c}|\widetilde{B}, \phi] =c_{z} \phi(c).
\end{equation}
Here,
\begin{equation}
\label{a.3}
\widetilde{\zeta} \equiv \frac{\zeta (t) \ell}{v_{0}(t)}\, ,
\end{equation}
\begin{equation}
\label{a.4}
\widetilde{B}({\bm c}) \equiv \frac{v_{0}^{d}(t) B({\bm v})}{\ell} = \frac{\chi ({\bm c}) \phi (c)}{a \ell},
\end{equation}
and $\widetilde{J}_{B}$ is the dimensionless Boltzmann collision operator,
\begin{equation}
\label{a.5}
\widetilde{J}_{B}[{\bm c}|\widetilde{B},\phi] = \int d{\bm c}_{2} \int d\widehat{\bm \sigma}\, \theta ({\bm c}_{12} \cdot \widehat{\bm \sigma})
 {\bm c}_{12} \cdot \widehat{\bm \sigma} \left[ b_{\bm \sigma}^{-1}(1,2)-1 \right] B({\bm c}_{1})  \phi (c_{2}).
 \end{equation}
The operator $b_{\bm \sigma}^{-1}(1,2)$ acts on ${\bm c}_{1}$ and ${\bm c}_{2}$ in the same way as on ${\bm v}_{1}$ and ${\bm v}_{2}$.

The formal solution of Eq.\ (\ref{a.2}) can be written as
\begin{equation}
\label{a.6}
\widetilde{B} ({\bm c}) = -\int_{0}^{\infty} ds\, e^{s\Lambda ({\bm c})} c_{z} \phi(c),
\end{equation}
where $\Lambda ({\bm c})$ is the linear operator defined by
\begin{equation}
\label{a.7}
\Lambda ({\bm c}) h({\bm c}) \equiv g_{e}(n) \widetilde{J}_{B}[{\bm c}| h,\phi] -\frac{\widetilde{\zeta}}{2} \frac{\partial}{\partial {\bm c}}\, \cdot \left[ {\bm c} h({\bm c})\right],
\end{equation}
for an arbitrary function $h({\bm c})$. From Eqs.\ (\ref{3.16}), (\ref{a.4}), and (\ref{a.6}), it is seen that
\begin{eqnarray}
\label{a.8}
D &=& \ell v_{0}(t) \int_{0}^{\infty} ds\, \int d{\bm c}\ c_{z} e^{s \Lambda ({\bm c})} c_{z} \phi (c) \nonumber \\
&=& \ell v_{0}(t) \int_{0}^{\infty} ds\, \left[ e^{s \Lambda^{+} ({\bm c})} c_{z} \right] c_{z} \phi(c).
\end{eqnarray}
In the above expression, $\Lambda^{+}({\bm c})$ is the adjoint of ${\bm c}$ given by
\begin{equation}
\label{a.9}
\Lambda^{+}({\bm c}_{1}) h({\bm c}_{1}) = g_{e}(n) \int d{\bm c}_{2} \phi(c_{2}) \int d \widehat{\bm \sigma} \theta ({\bm c}_{12} \cdot \widehat{\bm \sigma}) {\bm c}_{12} \cdot \widehat{\bm \sigma} \left[ b_{\bm \sigma} (1,2)-1 \right] h({\bm c}_{1}) - \frac{\widetilde{\zeta}}{2} {\bm c}_{1} \cdot \frac{\partial}{\partial {\bm c}_{1}}\, h({\bm c}_{1}).
\end{equation}
The operator $b_{\bm \sigma}(1,2)$ is the inverse of $b_{\bm \sigma}^{-1}(1,2)$, i.e. it changes all the velocities ${\bm c}_{1}$ and ${\bm c}_{2}$ to its right into the post-collisional values ${\bm c}^{\prime}_{1}$ and ${\bm c}^{\prime}_{2}$ given by
\begin{equation}
\label{a.10}
{\bm c}^{\prime}_{1} ={\bm c}_{1}- \frac{1+\alpha}{2}\, {\bm c}_{12} \cdot \widehat{\bm \sigma} \widehat{\bm \sigma},
\end{equation}
\begin{equation}
\label{a.11}
{\bm c}^{\prime}_{2} ={\bm c}_{2}+ \frac{1+\alpha}{2}\, {\bm c}_{12} \cdot \widehat{\bm \sigma} \widehat{\bm \sigma}.
\end{equation}
Now the approximation is made of considering $c_{z}$ as an eigenfunction of the operator $\Lambda^{+}({\bm c})$, i.e.
\begin{equation}
\label{a.12}
\Lambda^{+}({\bm c}) c_{z} \approx \lambda c_{z}.
\end{equation}
Then, Eq. (\ref{a.8}) becomes
\begin{equation}
\label{a.13}
D(t) = -\frac{\ell v_{0}(t)}{2 \lambda},
\end{equation}
where it has been assumed that $\lambda<0$, something to be taken into account to check the consistency of the approximation. To determine $\lambda$, Eq.\ (\ref{a.12}) is multiplied by $c_{z} \phi (c)$ and afterwards integrated over ${\bm c}$. After some simple algebra, it is obtained that
\begin{equation}
\label{a.14}
\lambda = - \frac{(1+\alpha) \widetilde{\zeta}}{2 (1-\alpha)}\, .
\end{equation}
Therefore, the eigenvalue turns out to be negative as it should for consistency. In the calculations, the expression for $\phi$ given in Eq.\ (\ref{2.15}) has been used, and contributions quadratic in $a_{2}$ have been neglected. Substitution of Eq. (\ref{a.14}) into Eq.\ (\ref{a.13}) leads to Eq.\ (\ref{3.17}), after using the expression for the cooling rate given in Eq.\ (\ref{2.18}).

Due to the symmetry of the expression of the self-diffusion coefficient as given by Eq.\ (\ref{a.8}), it follows that the approximation introduced by Eq.\ (\ref{a.12}) is actually equivalent in the present case to
\begin{equation}
\label{a.15}
\Lambda ({\bm c}) c_{z} \phi (c) \approx \lambda c_{z} \phi (c).
\end{equation}
Use of this in Eq.\ (\ref{a.6}) leads to
\begin{equation}
\label{a.16}
\widetilde{B}({\bm c}) = - \frac{2D(t)}{\ell v_{0}(t)}\, c_{z} \phi (c),
\end{equation}
and then, Eq.\ (\ref{3.19}) follows directly.


\begin{thebibliography}{99}

\bibitem{Go03} I. Goldhirsch, ``Rapid Granular Flows,'' Ann. Rev. Fluid Mech. {\bf 35}, 267 (2003).

\bibitem{ByP04} N.V. Brilliantov and T. P\"{o}schel, {\em Kinetic Theory of Granular Gases} (Oxford University Press, Oxford, 2004).

\bibitem{Go99} I. Goldhirsch, ``Granular gases: probing the boundaries of hydrodynamics,'' in {\em Granular Gases}, edited by T. P\"{o}schel and S. Luding (Springer-Verlag, Berlin, 1999).

\bibitem{DyvB77} J.R. Dorfman and H. van Beijeren, ``The Kinetic Theory of Gases,'' in {\em Statistical Mechanics, Pt. B}, edited by B.J. Berne (Plenum Press, New York, 1977).

\bibitem{McL89} J.A. McLennan, {\em Introduction to non-equilibrium statistical mechanics} (Prentice Hall, Englewoods Cliffs, New Jersey, 1989).

\bibitem{BRCyG00} J.J. Brey, M.J. Ruiz-Montero, D. Cubero, and R. Garc\'{\i}a-Rojo, ``Self-diffusion in freely evolving granular gases,'' Phys. Fluids {\bf 12}, 876 (2000).

\bibitem{ByP00} N.V. Brilliantov and T. P\"{o}schel, ``Self-diffusion in granular gases,'' Phys. Rev. E {\bf 61}, 1716 (2000).

\bibitem{DByL02} J.W. Dufty, J.J. Brey, and J. Lutsko, ``Diffusion in a granular fluid. I. Theory,'' Phys. Rev. E {\bf 65}, 051303 (2002).

\bibitem{LByD02} J. Lutsko, J.J. Brey, and J.W. Dufty, ``Diffusion in a granular fluid. II. Simulations,'' Phys. Rev. E {\bf 65}, 051304 (2002).



\bibitem{ByR13a} J.J. Brey and M.J. Ruiz-Montero, ``Steady self-diffusion in classical gases,''  EPL {\bf 103}, 30010 (2013).

\bibitem{GyZ93} I. Goldhirsch and G. Zanetti, ``Clustering instability in dissipative gases,'' Phys. Rev. Lett. {\bf 70}, 1619 (1993).

\bibitem{BRyM98} J.J. Brey, M.J. Ruiz-Montero, and F. Moreno, ``Instability and spatial correlations in a dilute granular gas,'' Phys. Fluids {\bf 10}, 2976 (1998).

\bibitem{BDyS97} J.J. Brey, J.W. Dufty, and A. Santos, ``Dissipative dynamics for hard spheres,'' J. Stat. Phys. {\bf 87}, 1051 (1997).

\bibitem{vByE79} H. van Beijeren and M.H. Ernst,  ``Kinetic theory of hard spheres,''   J. Stat. Phys. {\bf 21}, 125 (1979).

\bibitem{GyS95} A. Goldshtein and M. Shapiro, ``Mechanics of collisional motion of granular materials. 1. General hydrodynamic equations,'' J. Fluid Mech. {\bf 282}, 75 (1995).

\bibitem{vNyE98}T.P.C. van Noije and M.H. Ernst, ``Velocity distributions in homogeneous granular fluids: the free and the heated case,'' Granular Matter {\bf 1}, 57 (1998).



\bibitem{RydL77} P. R\'{e}sibois and M. de Leener, {\em Classical Kinetic Theory of Fluids} (Wiley, New York, 1977).




\bibitem{BMyG12} J.J. Brey, P. Maynar, and M.I. Garc\'{\i}a de Soria, ``Fluctuating Navier-Stokes equations for inelastic hard spheres or disks,'' Phys. Rev E {\bf 83}, 041303 (2011).

\bibitem{BRyC96} J.J. Brey, M.J. Ruiz-Montero, and D. Cubero, ``Homogeneous cooling state of a low-density granular flow,'' Phys. Rev. E {\bf 54}, 3664 (1996).

\bibitem{ByP06} N.V. Brilliantov and T. P\"{o}schel, ``Breakdown of the Sonine expansion of the velocity distribution of granular gases,'' Europhys. Lett. {\bf 74}, 424 (2006).

\bibitem{Bi94} G.A. Bird, {\em Molecular Gas Dynamics and the Direct Simulation of Gas Flows} (Clarendon, Oxford, 1994).

\bibitem{Ga00} A particularly clear and useful reference is A. L. Garc\'{\i}a, {\em Numerical methods for Physics} (Prentice Hall, Englewood Hills, NJ, 2000).


\bibitem{LyE72} A.W. Lees and S.F. Edwards, ``Computer study of transport processes under extreme conditions,'' J. Phys. C: Solid State Phys. {\bf 5}, 1921 (1972).

\bibitem{EyW77} J.J. Erpenbeck and W.W. Wood, ``Molecular dynamics techniques for hard-core systems,'' in {\em Statistical Mechanics, Pt. B}, edited by B.J. Berne (Plenum Press, New York, 1977).



\bibitem{Lu01} J.F. Lutsko, ``Model for the atomic-scale structure of the homogeneous cooling state of granular fluids,'' Phys. Rev. E {\bf 63}, 061211 (2001).

\bibitem{BRyM04} J.J. Brey, M.J. Ruiz-Montero, and F. Moreno, ``Steady state representation of the homogeneous cooling state of a granular gas,'' Phys. Rev. E {\bf 69}, 051303 (2004).


\end{thebibliography}
\end{document}